\documentclass[conference,a4paper]{IEEEtran}
\pdfoutput=1

\usepackage{amsfonts}
\usepackage{amsmath}
\usepackage{amssymb}
\usepackage{amsthm}
\usepackage{balance}
\usepackage{booktabs}
\usepackage{cite}
\usepackage{graphicx}
\usepackage{tikz}
\usepackage{xspace}
\usepackage[T1]{fontenc}
\usepackage[colorinlistoftodos]{todonotes}
\usepackage[english]{babel}
\PassOptionsToPackage{hyphens}{url}\usepackage[hidelinks]{hyperref}
\usepackage[utf8x]{inputenc}

\usepackage[a4paper,top=1.9cm,bottom=4.3cm,left=1.3cm,right=1.3cm]{geometry}

\newcommand*\circled[1]{\tikz[baseline=(char.base)]{
            \node[shape=circle,draw,inner sep=1pt] (char) {#1};}}

\newcommand{\lmst}{\fontfamily{lmtt}\selectfont} 

\newcommand{\para}[1]{\vspace{0.08in}\noindent\textbf{#1 }}

\newcommand{\name}{{\normalfont\lmst AGasP}\xspace}

\begin{document}
\title{Smart Contracts for Machine-to-Machine Communication: Possibilities and Limitations}

\author{
  \IEEEauthorblockN{Yuichi Hanada} \\
  \IEEEauthorblockA{Stanford University\\
                    yuhanada@stanford.edu}
\and
  \IEEEauthorblockN{Luke Hsiao} \\
  \IEEEauthorblockA{Stanford University\\
                    lwhsiao@stanford.edu}
\and
  \IEEEauthorblockN{Philip Levis} \\
  \IEEEauthorblockA{Stanford University\\
                    pal@cs.stanford.edu}
}

\newcommand\copyrighttext{%
  \footnotesize © 2018 IEEE. Personal use of this material is permitted.
  Permission from IEEE must be obtained for all other uses, in any current or
  future media, including reprinting/republishing this material for advertising
  or promotional purposes, creating new collective works, for resale or
  redistribution to servers or lists, or reuse of any copyrighted component of
  this work in other works.
  DOI: \href{https://ieeexplore.ieee.org/document/8600854}{10.1109/IOTAIS.2018.8600854}}
  \newcommand\copyrightnotice{%
  \begin{tikzpicture}[remember picture,overlay]
  \node[anchor=south,yshift=10pt] at (current page.south) {\fbox{\parbox{\dimexpr\textwidth-\fboxsep-\fboxrule\relax}{\copyrighttext}}};
  \end{tikzpicture}%
}

\maketitle
\copyrightnotice
\begin{abstract}
Blockchain technologies, such as smart contracts, present a unique interface for
machine-to-machine communication that provides a secure, append-only record that
can be shared without trust and without a central administrator. We study the
possibilities and limitations of using smart contracts for machine-to-machine
communication by designing, implementing, and evaluating \name, an application
for automated gasoline purchases. We find that using smart contracts allows us
to directly address the challenges of transparency, longevity, and trust in IoT
applications. However, real-world applications using smart contracts must
address their important trade-offs, such as performance, privacy, and the
challenge of ensuring they are written correctly.
\end{abstract}


\section{Introduction}
\label{sec:intro}

The Internet of Things (IoT) refers broadly to interconnected devices that
communicate, share data, measure the physical world, and interact with people.
IoT applications have been deployed in a wide variety of domains such as
healthcare, manufacturing, agriculture, and transportation. They also have the
potential to transform daily life through smart homes, cities, and
infrastructure~\cite{zanella2014internet, soliman2013smart, zhang2012cognitive}.
Cisco and Ericsson estimate that there will be 100 billion IoT devices by
2020~\cite{weyrich2014machine}.

\begin{figure}[t]
\centering
\includegraphics[width=0.60\columnwidth]{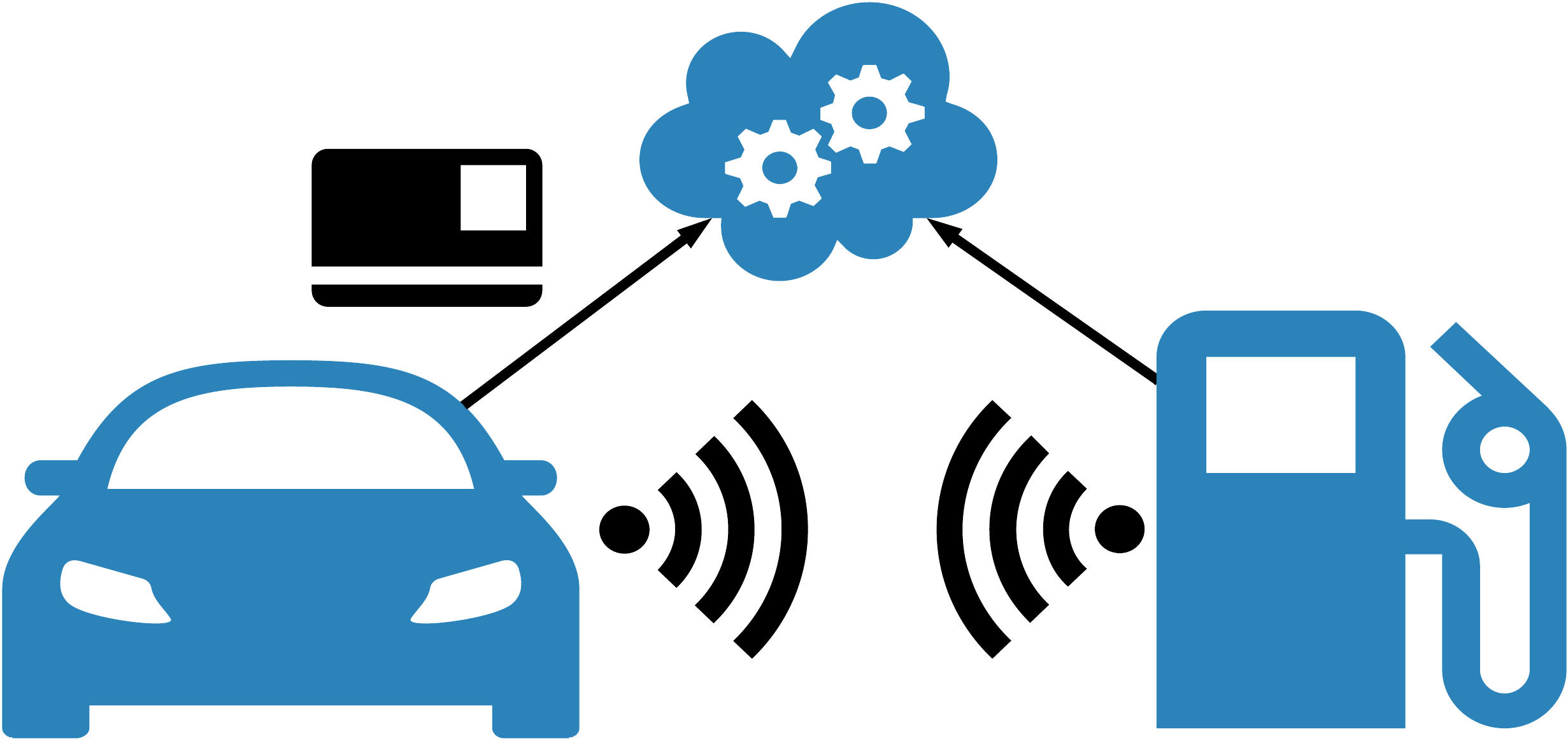}
\caption{A traditional IoT application that stores a user's credit card
information and is installed in a smart vehicle and smart gasoline pump. Before
refueling, the vehicle and pump communicate directly using short-range wireless
communication, such as Bluetooth, to identify the vehicle and pump involved in
the transaction. Then, the credit card stored by the cloud service is charged
after the user refuels. Note that each piece of the application is controlled
by a single entity.}
\label{fig:intro}
\vspace{-1em}
\end{figure}

Many IoT applications rely on \emph{machine-to-machine communication}---the
communication between devices with limited or without human
intervention~\cite{chen2012machine}---to automate tasks, send commands, and/or
distribute information. Figure~\ref{fig:intro} shows an example of this broad
class of machine-to-machine applications, a smart vehicle automatically paying
for its refueling. Machine-to-machine applications encounter three technical
challenges that human-centric applications solve with a person in the loop.

The first challenge is transparency. IoT devices are infamously
insecure~\cite{grover2016internet, enterprise2015internet}, but encryption makes
it extremely difficult to verify that they are acting appropriately. In
Figure~\ref{fig:intro}, where actions are automatically performed by a smart
vehicle on a user's behalf, it would be impossible for the user to audit the
encrypted information sent from their vehicle to the cloud service to ensure
that private data was not also sent---a concern that is well
justified~\cite{angwin2015vizio, oltermann2017doll}.

The second challenge is longevity. IoT devices are often expected to function
for decades. Some of these objects may still be in use long after their vendors
stop maintaining them~\cite{hartzog2015internet}. But, because these devices are
often \emph{vertical silos}---where a centralized entity manages application
state and communication protocols---they cannot function without the cloud
services of their vendors~\cite{hern2016revolv, gallagher2016printer}. In
Figure~\ref{fig:intro}, without the cloud service, the vehicle is unable to pay
the gas station.

The third challenge is trust. In Figure~\ref{fig:intro}, the user must trust the
vendor with their credit card information, and the credit card company acts as a
trusted third party to help manage funds and resolve disputes in exchange for
non-negligible fees. IoT transactions that involve the exchange of digital or
physical assets require trust, which inherently involves risks. Examples of such
risks are the vendor leaking credit card information or the credit card company
undermining the fairness of an exchange by colluding when resolving
disputes~\cite{goldfeder2017escrow}. Each of these risks can compromise an IoT
application.

To summarize these challenges:
\begin{enumerate}
\item IoT applications acting on behalf of users benefit from
  \emph{transparency}. Without it, users cannot verify the actions their
  applications are taking on their behalf.
\item Long-lived IoT devices may outlive the infrastructure that supports them,
  causing them to fail or exposing vulnerabilities. They lack \emph{longevity}.
\item IoT applications that exchange goods or services require \emph{trust}.
  This often adds financial overhead and risks that can subvert an application.
\end{enumerate}

One approach to addressing these challenges is to use blockchain technologies. A
blockchain provides a publicly-auditable, append-only ledger that ensures full
transparency of transactions performed on the chain. It also allows state to be
stored in a distributed manner among the nodes in a network that persists as
long as a network of nodes exists. Furthermore, blockchain-based ledgers that
support smart contracts allow applications to embed the contractual logic of a
transaction onto the blockchain. This way, the logic is executed independently
and automatically by each node on the network using the data provided on the
blockchain---reducing the need for trust and third party involvement in a
transaction~\cite{szabo1997formalizing}.

However, these beneficial properties come with impactful trade-offs: certain
blockchain consensus algorithms, such as the proof-of-work used in Ethereum,
significantly limit the performance of executing transactions in terms of both
throughput and latency; the availability of a public log of transactions raises
important privacy considerations; and smart contracts require special care in
deployment. Because a smart contract acts as the absolute arbiter for
transactions performed through it, bugs in a smart contract may incur costly
damages that are difficult or impossible to
resolve~\cite{siegel2016understanding}.

Researchers have proposed a variety of uses for blockchain and smart contracts
in IoT applications (Section~\ref{sec:related}). Our work presents an initial
technical and experimental evaluation of this approach. We present \name, an IoT
application for automated, machine-to-machine gasoline purchases that uses smart
contracts to perform transactions. In contrast to the traditional centralized
IoT architecture shown in Figure~\ref{fig:intro}, \name allows users to audit
all interactions and can continue to function as long as the blockchain network
exists. Furthermore, by using smart contracts to cut out third parties, \name
can save gas stations over 79\% in fees for a typical transaction.

\para{Contributions} This paper makes these contributions:
\begin{enumerate}
\item The design, implementation, and evaluation of \name, an application that
  uses smart contracts for automated gasoline purchases between a vehicle and
  gas station.
\item A discussion of the practical trade-offs of using smart contracts for
  machine-to-machine communication. We find that transaction latencies are
  sensitive to transaction fees and that the 95-percentile latency can
  be on the order of hours (Section~\ref{sec:limitations}).
\end{enumerate}

The remainder of the paper is structured as follows.
Section~\ref{sec:background} describes the building blocks for using smart
contracts. Section~\ref{sec:case} makes the case for smart contracts.
Section~\ref{sec:system} describes \name's design and implementation while
Section~\ref{sec:eval} evaluates its performance and
Section~\ref{sec:limitations} discusses the limitations of this approach. We
conclude in Section~\ref{sec:conclusion}.

\section{Background}
\label{sec:background}

We describe blockchains and smart contracts abstractly, and then describe how
they are implemented in Ethereum, the largest smart contract blockchain used
today, to ground these abstract concepts concretely.

\subsection{Blockchain}
\label{sec:blockchain}

A \emph{blockchain} is a distributed data structure, introduced with
Bitcoin~\cite{nakamoto2008bitcoin}, that provides a verifiable, append-only
ledger of transactions. As its name suggests, a blockchain is comprised of
timestamped blocks, where each block is identified by a cryptographic hash and
references the hash of the block preceding it---forming the links, or chain,
between each block. In addition, blocks may contain \emph{transactions}, which
record a transfer of data or assets between two addresses. As a result, any node
in a \emph{blockchain network} with access to the blockchain can traverse it to
construct the global state stored on the chain. Nodes in a blockchain network
operate on the same blockchain and form a peer-to-peer network where each node
replicates all or part of the blockchain.

In order to submit transactions to the chain, each node uses a pair of public
and private keys. First, the node constructs and signs a transaction and
broadcasts it to its one-hop peers. Each node validates any transactions it
receives, dropping invalid transactions, before broadcasting it to its peers.
These transactions form a pool of valid, pending transactions that are ready to
be included in a block. \emph{Miners} are nodes in the network that construct
new blocks to be added to the blockchain and broadcasts them to its peers, who
verify it before appending it. This process continuously repeats. Each
blockchain employs a consensus mechanism to resolve different states, or
``forks'' in the network, and the choice of mechanism varies among networks.
\cite{tschorsch2016bitcoin} details a variety of consensus mechanisms.

\subsection{Smart Contracts}
\label{sec:scs}

\emph{Smart contracts} ``combine protocols, user interfaces, and promises
expressed via those interfaces, to formalize and secure relationships over
public networks~\cite{szabo1997formalizing}.'' In other words, smart contracts
allow users to execute a script on a blockchain network in a verifiable way and
allows many problems to be solved in a way that minimizes the need for trust. To
do so, they allow users to place trust directly in the deterministic protocols
and promises specified in a smart contract, rather than in a third party. For
example, in Figure~\ref{fig:intro}, users must trust the vendor with their
credit card number, and the gas station must trust the credit card company to
pay on behalf of the user in exchange for fuel. With smart contracts, a user can
pay the gas station directly by using the protocol and promises established in
the contract such that neither party can manipulate the exchange.

A smart contract has its own address and account on the blockchain.
Consequently, it can maintain its own state and take ownership of assets on the
blockchain, which allows it to act as an escrow. Smart contracts expose an
interface of functions to the network that can be triggered by sending
transactions to the smart contract. Because a smart contract resides on the
blockchain, each node can view and execute its instructions, as well as see the
log of each interaction with each smart contract. A smart contract acts as an
autonomous entity on the blockchain that can deterministically execute logic
expressed as functions of the data that is provided to it on the blockchain.

\subsection{The Ethereum Platform}
\label{sec:ethereum}

Ethereum is an open-source, distributed platform based on a blockchain. In
Ethereum's blockchain network, miners execute a \emph{proof-of-work} consensus
algorithm~\cite{wood2014ethereum}. In addition, Ethereum supports smart
contracts written in a Turing-complete language, like Solidity, which can be
compiled to bytecode that can be executed in the Ethereum Virtual Machine. This
allows users to create arbitrary ownership rules, transaction formats, and state
transition functions~\cite{buterin2014next}.

In Ethereum, all computation and transactions have fees, which are measured in
units of \texttt{gas} (not to be confused with the real-world fuel discussed in
our running example). Each transaction in Ethereum must specify a
\texttt{gasLimit}, which is the maximum amount of \texttt{gas} that may be used
while executing a transaction. Transactions also specify a \texttt{gasPrice},
which is the rate paid to miners in Ether (Ethereum's associated cryptocurrency)
per unit of \texttt{gas} as a reward. If an account cannot support the maximum
fee of $(\texttt{gasPrice} * \texttt{gasLimit})$, the transaction is considered
invalid. The amount of \texttt{gas} used (\texttt{gasUsed}) is determined by the
amount of computation and storage required by a transaction. A transaction's fee
is calculated as shown in Equation~\ref{eq:fee}.
\begin{equation}
\label{eq:fee}
Fee = \texttt{gasPrice} * \min(\texttt{gasUsed}, \texttt{gasLimit})
\end{equation}
Transactors may specify any positive \texttt{gasPrice} and \texttt{gasLimit}.
Likewise, miners may ignore transactions and prioritize transactions with larger
fees to maximize their profits. In Ethereum, the maximum number of transactions
that can be included in a block is limited by the total amount of \texttt{gas}
used by the transactions in the block. Unlike Bitcoin's size-based limit of 1MB,
this \texttt{gas} block limit can increase by a small scalability factor each
block~\cite{wood2014ethereum}. These block limits create an upper-bound for
transaction throughput of the network if the mining rate of new blocks remains
constant.

Relaxing the block size limit to improve throughput is not a simple solution.
For example, increasing the block size increases the resources required to run a
full mining node, which could lead to centralization of entities with high
compute power. While other approaches for improving scalability such as
off-chain solutions (e.g., the Raiden Network~\cite{raiden}), periodic
merkleized commits~\cite{poon2017plasma}, and sharding~\cite{sharding} have been
proposed, we only consider the implementation of the current Ethereum network
for this work.

\subsection{Related Work}
\label{sec:related}

Many researchers have explored the application of smart contracts in the IoT
domain. For example, smart contracts have been proposed as a mechanism for
managing access control~\cite{zhang2018smart, xu2018blendcac,
Novo2018BlockchainMI} and authentication~\cite{lee2018implementation}. Others
have investigated security and privacy implications of using smart contracts in
IoT applications~\cite{dorri2017blockchain}, explored using smart contracts to
create a shared marketplace of services between
devices~\cite{christidis2016blockchains, huckle2016internet}, and built
industrial IoT platforms using blockchains~\cite{bahga2016bpiiot}.

\section{The Case for Smart Contracts}
\label{sec:case}

We find that smart contracts address the technical challenges of transparency,
longevity, and trust that are frequently encountered in IoT applications like
Figure~\ref{fig:intro}.

\para{Transparency with Public Logs}
In Figure~\ref{fig:intro}, the vendor may wish to gather personal information
about a user (such as driving behavior, location history, or vehicle mileage)
that could then be sold to advertisers for better advertisement targeting. Users
would have no way to directly verify exactly what information was being sent by
their vehicle if communication to the cloud service was encrypted. The use of
smart contracts as the interface for machine-to-machine communication provides a
public, auditable log of communication.

\para{Longevity through Decentralization}
In 2016, the average age of a car in the United States was 11.6
years~\cite{ihs2016cars}---a time frame that may outlast the vendor of our
example IoT application. By using smart contracts, the core state and logic of
an application is fully distributed, allowing an application to continue to
operate or be picked up by a new vendor long after the original vendor has shut
down. Because smart contracts are a public interface, anyone can directly
interact with the smart contract using their own applications with the assurance
that the application's state will be available on the blockchain, rather than
being lost with the shutdown of a vendor.

\para{Minimized Trust using Smart Contracts}
The need to trust vendors or third parties with personal information like credit
cards or bank accounts creates valuable targets for attack, many of which have
been notoriously exploited~\cite{o2017giant}. Instead, transactors can avoid
this risk by interacting directly through deterministic business logic specified
in smart contracts and using smart contracts as reliable escrows for digital
assets, creating a platform that supports a wide variety of applications.

\section{\name: Automated Gasoline Purchases}
\label{sec:system}

\begin{figure}[t]
\centering
\includegraphics[width=0.60\columnwidth]{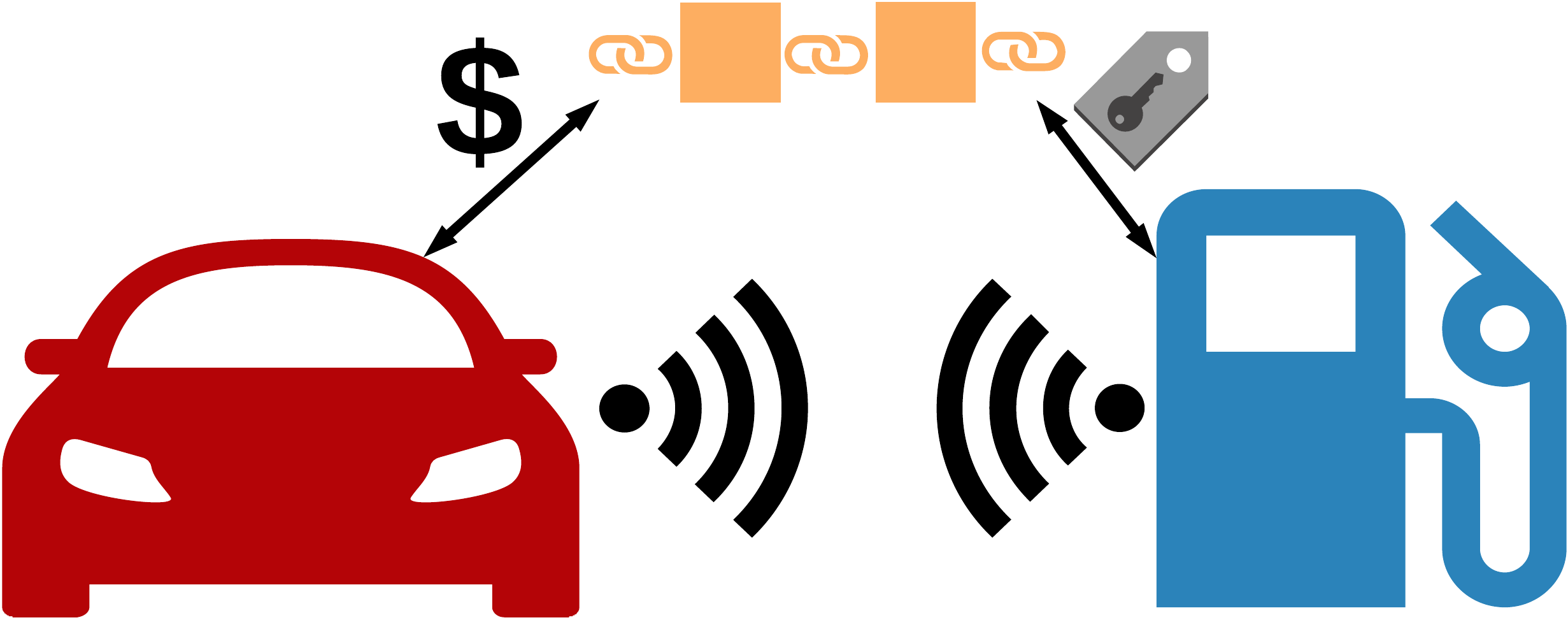}
\caption{\name uses smart contracts in Ethereum to pay for fuel. There is no
centralized entity that manages the state of the application. Instead, a
vehicle running a decentralized application can interact directly with the
public blockchain to submit funds to the smart contract. Likewise, the gas
station can interact directly with the blockchain through its application to
determine whether a vehicle has paid, and to record how much gas was
purchased.}
\label{fig:agasp_overview}
\vspace{-1em}
\end{figure}

To understand the trade-offs involved with using smart contracts, we design,
implement, and evaluate \name, a decentralized application for automated
gasoline purchases. In contrast to the centralized approach of
Figure~\ref{fig:intro}, with \name, the state and protocol of the application
are not controlled by a single vendor, as shown in
Figure~\ref{fig:agasp_overview}. Instead, smart contracts on the Ethereum
blockchain network are used to purchase gasoline in a verifiable and auditable
way that minimizes trust.

\subsection{System Design}
\label{sec:overview}

In \name, the smart contract is the main component since it defines the protocol
of a gas purchase in a general way, and stores the state and logic necessary to
complete a purchase. Consequently, a single contract can be utilized by many
different vehicles and different gas stations---it serves as a common interface
and protocol for purchasing gasoline. We designed the \name smart contract
protocol to follow a sequence familiar to users from traditional gasoline
purchases, while removing both the credit card company and vendor as third
parties.

When exchanging digital currency, smart contracts themselves can act as an
escrow to ensure that either both parties get the results of the expected
exchange, or neither do (e.g., if one party were to cancel). However, in the
case of exchanging a physical good, the smart contract cannot act as an escrow.

Instead, the smart contract must clearly specify the sequence of events in order
to minimize the risk involved in the exchange. For example, once gasoline is
dispensed, it cannot be returned. In order to eliminate the risk of a vehicle
refueling and trying to avoid payment, we require that payment occur before a
vehicle is allowed to refuel, following the pattern of traditional gasoline
payment. Similarly, our contract protects gas stations from malicious users by
placing the control of completing an exchange in the power of gas stations
(i.e., a vehicle cannot refuel a large amount and then claim only a small amount
was taken). Finally, the smart contract itself contains all of the information
necessary to calculate the payment to send to the gas station, and the change to
return to the vehicle. With this design, the vehicle cannot withdraw its deposit
without the gas station's involvement in order to protect against the case where
a vehicle attempts to withdraw a deposit while refueling. This places trust in
the gas station to report the correct amount of fuel dispensed, as is
traditionally done.

\begin{figure}[t]
\centering
\includegraphics[width=1.0\columnwidth]{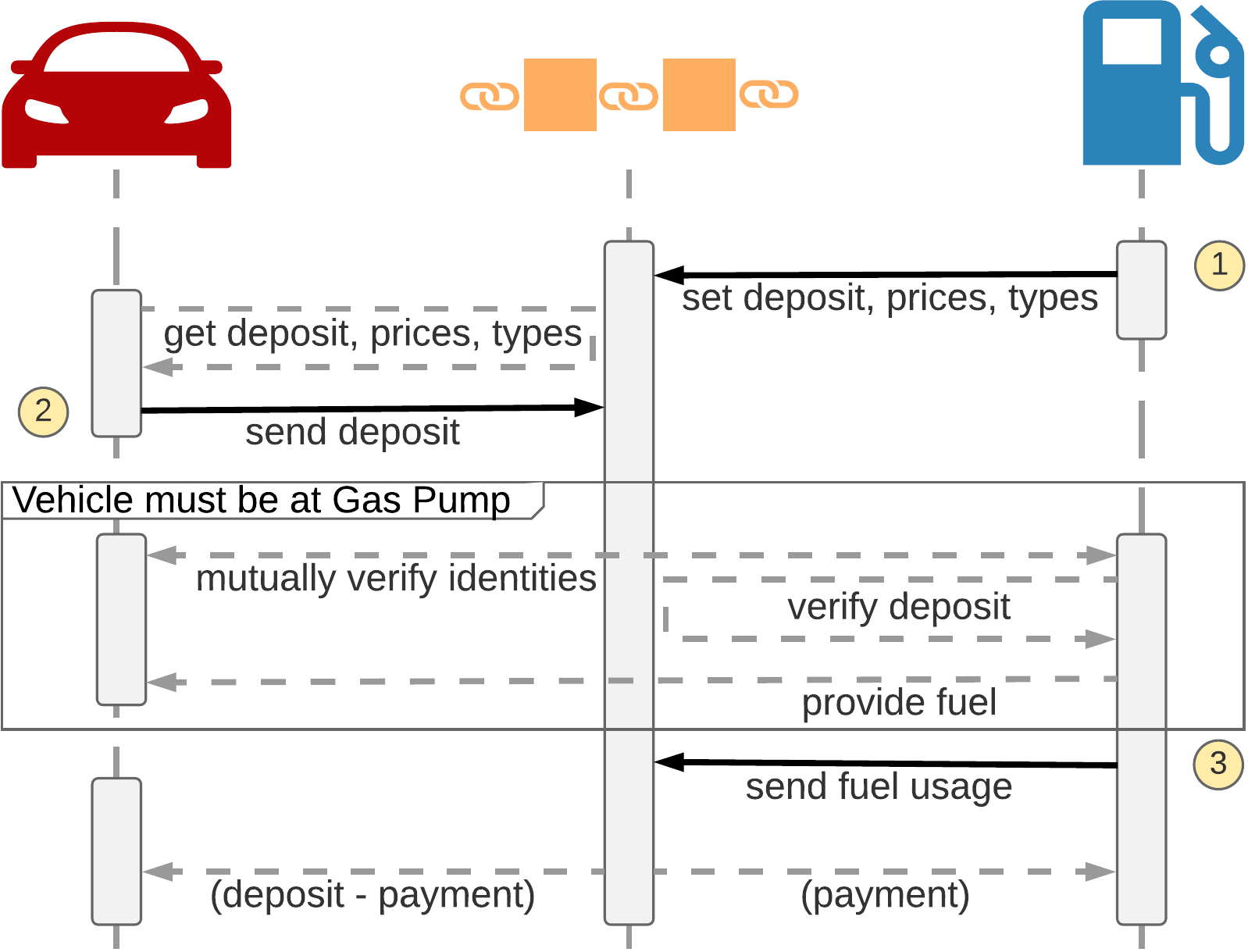}
\caption{A typical transaction sequence between a vehicle, the \name smart
contract, and a gas station during a gasoline purchase. Ethereum transactions
that require the payment of fees are drawn with a solid black line.}
\label{fig:sequence}
\vspace{-1em}
\end{figure}

In \name, a smart contract for gasoline purchases is published to the
blockchain. The protocol begins with a gas station publishing a minimum deposit
amount, as well as prices for their various types of gasoline. A user then
transfers Ether to their vehicle. When the user decides that they would like to
refuel, they use the vehicle application to initiate a deposit to the gas
station where they plan to refuel. The gas station is then able to view this
transaction on the blockchain to identify which vehicles are authorized to fuel
at the station. When the vehicle arrives at the pump, a short-range wireless
protocol (e.g., Bluetooth) can be used to verify the identity of the vehicle and
the gas station, and the fueling can begin. Once fueling is complete, the
station sends a transaction indicating the amount and type of fuel dispensed and
the smart contract calculates the payment for the gas station, returning the
deposit to the vehicle after subtracting the payment (see
Figure~\ref{fig:sequence}).

\subsection{\name Implementation}
\label{sec:implementation}

To fulfill this protocol, the smart contract stores an internal list of gas
stations, along with the types of fuel they sell and their respective prices.
New stations are added to this list, and prices are adjusted in this list, by
calling \texttt{setGasInfo}, the function to set the minimum deposit, prices,
and types. A vehicle can poll the blockchain for station, type, and price
information by calling \texttt{getGasInfo}. When a vehicle calls
\texttt{sendDeposit} to send a deposit, it includes the address of the station
so that the smart contract can keep a list of vehicles and their deposits for
each station, along with the prices at the time of deposit. After verifying
identities, the station can check this list to ensure that a deposit has been
made before dispensing fuel by calling \texttt{verifyDeposit}. Once refueling is
complete and the amount of fuel dispensed is reported by calling
\texttt{sendFuelUsage}, the payment and change are sent, and the vehicle is
removed from the list. We implement the \name smart contract using Solidity, a
high-level language for the Ethereum platform.

In order to interact with the \name smart contract, we also developed two
decentralized applications---one for the vehicle and one for the gas
station---using JavaScript and the web3 Ethereum JavaScript API\@. Each
application is programmed with the address of the \name smart contract. The gas
station application provides an interface to call \texttt{setGasInfo},
\texttt{verifyDeposit}, and \texttt{sendFuelUsage}. The vehicle application
provides an interface for calling \texttt{getGasInfo} and \texttt{sendDeposit}.
These applications simply help us test and develop the \name smart contract.

\section{Evaluation}
\label{sec:eval}
We evaluate \name by seeking to answer: (1) does \name address the challenges
resulting from traditional approaches and (2) how does \name compare to
traditional approaches in terms of financial overhead to gas stations?

\subsection{\name Compared to a Centralized Approach}

\para{Transparency}
As shown in Figure~\ref{fig:sequence}, the only operations that do not interact
with the blockchain are the mutual verification of identities and the refueling.
The rest of the key components of a purchase are committed as blockchain
transactions. With this scheme, it is impossible for a vehicle to purchase
gasoline without making an auditable deposit on the chain to a specific gas
station, and it is impossible for a gas station to acquire their payment without
publishing the amount of fuel that was dispensed. Similarly, it is impossible
for a gas station to charge a price other than what was committed to the
blockchain at the time of purchase, since the smart contract performs the
calculation based on the published prices.

\para{Longevity}
In an architecture like Figure~\ref{fig:intro}, the application relies on the
vendor's infrastructure and cloud services to operate. With \name, the
infrastructure is completely distributed; the permanence of the infrastructure
and state of the application is based on the permanence of the Ethereum network.
Although Ethereum is still relatively nascent, because it is open-source, anyone
can set up new Ethereum nodes to ensure that applications persist. There are
threats to the longevity of using a public, proof-of-work-based blockchain like
Ethereum, such as a malicious entity controlling 51\% of a network's mining
power, allowing them to manipulate the ledger. However, the high cost of such an
attack on large public networks makes it unlikely~\cite{cawrey2014attack}, and
moving away from proof-of-work is already on the Ethereum
roadmap~\cite{hertig2017pos}.

\begin{figure}[t]
\centering
\includegraphics[width=0.7\columnwidth]{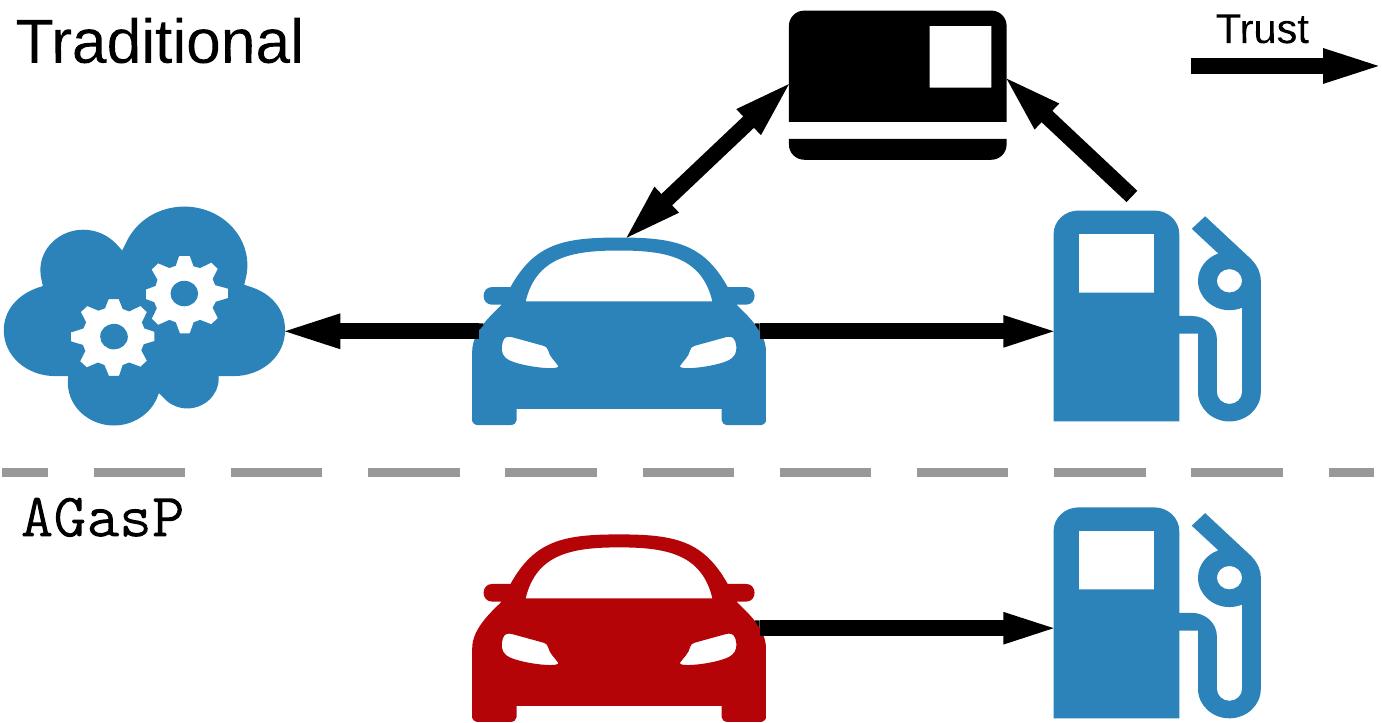}
\caption{Smart contracts allow applications to minimize trust.}
\label{fig:trust}
\vspace{-1em}
\end{figure}

\para{Trust}
In Figure~\ref{fig:intro}, there are several relationships of trust leveraged to
complete a transaction. The user must trust: the vendor with their credit card,
the credit card to pay the station, and the station to give them fuel and charge
the correct amount. Next, the credit card company trusts the user to pay back
their debt. Finally, the station trusts the credit card company to pay on behalf
of the user. By leveraging a smart contract, we can reduce the trust to just a
single edge (see Figure~\ref{fig:trust}): a user must trust the gas station to
give them fuel and charge the correct amount\footnote{While multi-signature
transactions can be used to distribute the trust among several parties (i.e.,
requiring $m$ of $n$ participants to validate a transaction), ultimately, at
least one point of trust is required when exchanging physical goods or
services~\cite{pagnia1999impossibility}.}. The smart contract itself will then
distribute payment without the involvement of another party.

\subsection{Transaction Fees}
\label{sec:fees}

In Figure~\ref{fig:intro}, the gas station must pay fees to the credit card
company, and many gas stations have passed those fees onto users by charging
more for credit card transactions than for debit or cash~\cite{simon2008gas}.
Similarly, with \name, both the user and the gas station must pay transaction
fees in order for their transactions to be included on the blockchain. In order
to estimate these savings, we estimate the amount of fees paid for a refueling a
small car. Credit card companies charge a fee comprised of a rate ($Rate$) of
the transaction amount ($Amount$) along with a flat fee ($Flat$) for each
transaction, as shown in Equation~\ref{eq:credit}.
\begin{equation}
\label{eq:credit}
Fee = Rate * Amount + Flat
\end{equation}
Assuming a typical rate of 2\% (less than the 2016
average~\cite{nilson2017merchant}), a transaction amount of filling a 12 gallon
tank with gas at \$3.85 per gallon, and a flat fee of \$0.25, we estimate a fee
of \$1.17. Similarly, we can compute an estimate of the transaction fees
incurred by \name using Equation~\ref{eq:fee}. If we assume a \texttt{gasPrice}
of $10\times10^{-9}$ Ether per transaction, and an Ether value of \$650. Then,
based on our experimental measurements of \texttt{gasUsed} for transactions
\circled{1}, \circled{2} and \circled{3} (as labeled in
Figure~\ref{fig:sequence}), the total cost in transaction fees for \name would
be \$0.78, a 33\% reduction in transaction fees. Furthermore, note that a gas
station only needs to send \circled{1} when gas prices change, and \circled{2}
is paid for by the vehicle. Thus, for a typical refueling, the gas station only
pays \$0.25---a 79\% reduction in transaction fees. These estimates are
summarized in Table~\ref{tab:fees}.

However, unlike credit card fees controlled by a credit card company, the
transaction fees of operating on Ethereum vary with the market and can be
unpredictable. For example, Ether was valued at over \$1300 in January 2018, and
dipped back to under \$380 in April 2018---a span of just three
months~\cite{ether2018price}.

\begin{table}[t]
\centering
\caption{Estimates of transaction fees with and without \name.}
\label{tab:fees}
\begin{tabular}{llr}
  \toprule
  \textbf{Transaction}                  & \textbf{Paying Party} & \textbf{Approx. Fee (\$)} \\
  \midrule
  Credit Card                           & Gas Station          & 1.17 \\
  \midrule
  \circled{1} \texttt{setGasInfo}       & Gas Station          & 0.21 \\
  \circled{2} \texttt{sendDeposit}      & Vehicle              & 0.32 \\
  \circled{3} \texttt{sendFuelUsage}    & Gas Station          & \textbf{0.25} \\
  \circled{1}, \circled{2}, \circled{3} & Vehicle, Gas Station & 0.78 \\
  \bottomrule
\end{tabular}
\end{table}

\section{Limitations and Discussion}
\label{sec:limitations}

\subsection{Performance}
\label{sec:performance}

A noticeable limitation of using smart contracts on Ethereum for
machine-to-machine communication is that low transaction throughput results in
high latencies in the time that it takes to complete a transaction. In
Figure~\ref{fig:intro}, a cloud service using a distributed database such as
Google Spanner~\cite{corbett2013spanner} can perform tens of thousands of
transactions per second. In stark contrast, the public Ethereum network's peak
throughput was just short of 16 transactions per second~\cite{ether2018trans}.
Furthermore, because miners are incentivized to prioritize transactions with
higher rewards, transaction time can be highly influenced by the
\texttt{gasPrice} of the transaction.

\begin{figure}[t]
\centering
\includegraphics[width=1.0\columnwidth]{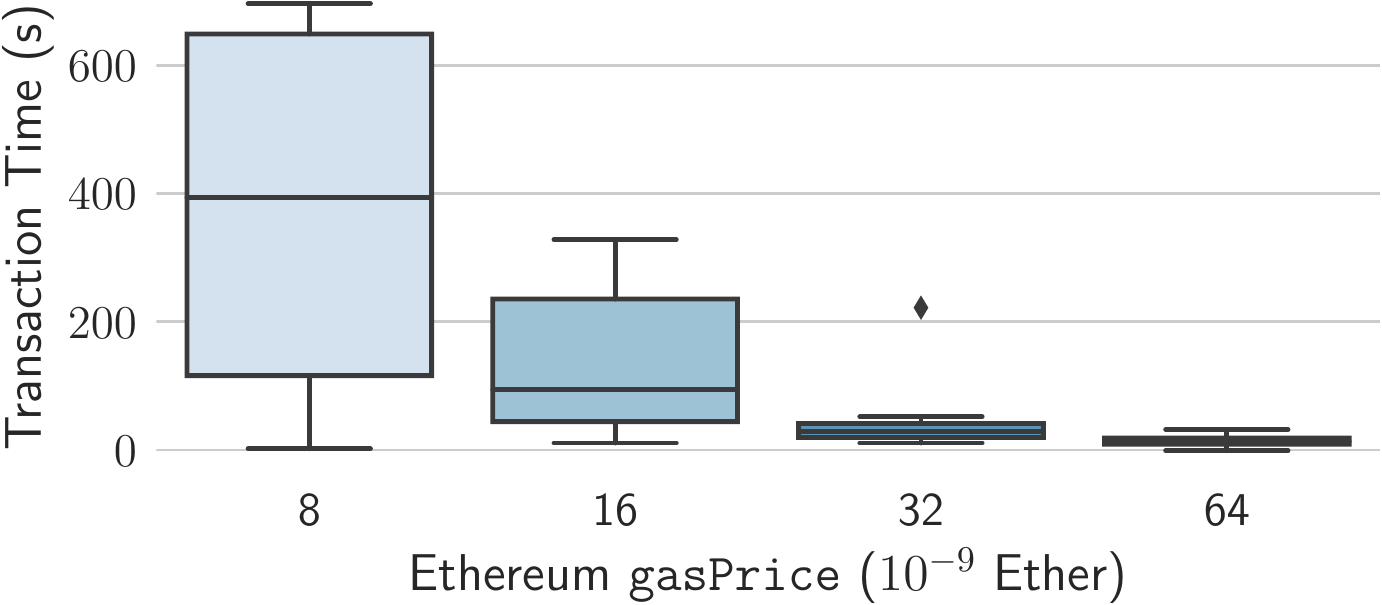}
\caption{Transaction times for increasing \texttt{gasPrice} on the public
Ethereum network with a \texttt{gasLimit} of 70,000.}
\label{fig:time_v_gas}
\end{figure}

\begin{figure}[t]
\centering
\includegraphics[width=1.0\columnwidth]{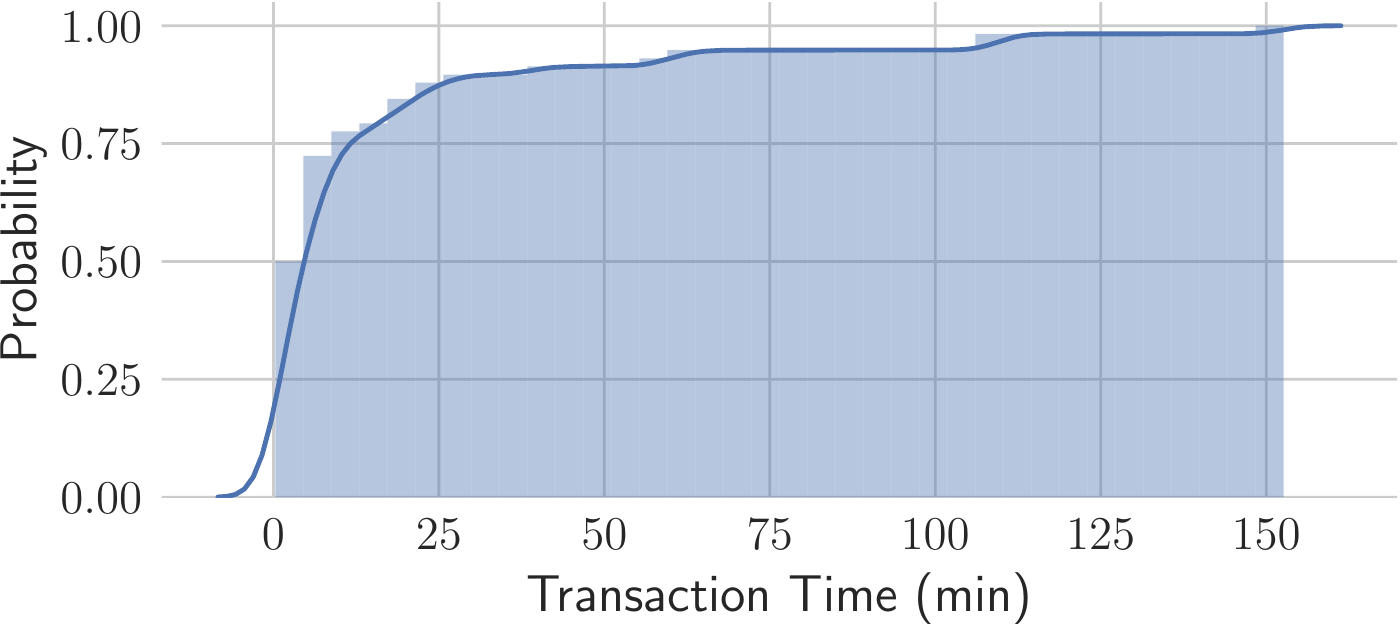}
\caption{Cumulative density function for the transaction times of
\texttt{setGasInfo} on the public Ethereum network with a \texttt{gasPrice} of
$10\times10^{-9}$ Ether and a \texttt{gasLimit} of 70,000.}
\label{fig:public_cdf}
\vspace{-1em}
\end{figure}

In Figure~\ref{fig:time_v_gas}, we vary the \texttt{gasPrice} of a transaction
while holding all other variables constant and measure the time it takes from
sending the transaction from a node on the public network, to seeing that
transaction executed and included in a block on the chain. Each
\texttt{gasPrice} was tested five times. A higher \texttt{gasPrice} helps reduce
both the average transaction time and the variance of transaction times, down to
an average of 14 seconds with a \texttt{gasPrice} of $64\times10^{-9}$ Ether.
However, we see diminishing returns as eventually we are bottlenecked by the
throughput of the Ethereum network. In Figure~\ref{fig:public_cdf}, we use a
constant \texttt{gasPrice} and \texttt{gasLimit}, and run the same transaction
60 times over the span of 5 days. The load of the network and value of
\texttt{gas} varies over time, leading to high tail latencies when using a
constant \texttt{gasPrice}. In our experiments, we measure a mean latency of 14
minutes and a 95-percentile latency of 69 minutes.

The relatively low performance of smart contracts on Ethereum means that
applications that utilize them must design around the possibility of transaction
delays on the order of minutes to hours, or be prepared to raise
\texttt{gasPrice} values to lower transaction times (down to the order of tens
of seconds). In \name, we design our sequence such that transaction latency can
be hidden from the user by allowing the time-consuming smart contract
transactions to occur before and after the user is interacting with the gas pump
(Figure~\ref{fig:sequence}). If a user does not pay beforehand, they must wait
one transaction time at the pump. Other blockchain networks, such as Stellar,
address performance by using alternative consensus
algorithms~\cite{mazieres2015stellar}.

\subsection{Privacy}
\label{sec:privacy}

Recall that transactions contain an origin address, a destination address, and
the data or assets to be sent. Although using smart contracts enables users to
audit the transactions of their devices, the nature of a public blockchain also
means that anyone else can also view these transactions. Even though addresses
are not explicitly tied to a real-world identity, other nodes are still able to
monitor a blockchain to learn patterns about a user's transactions.

Techniques such as \emph{zero-knowledge Succinct Non-interactive ARguments of
Knowledge} (zkSNARK)~\cite{ben2014succinct} have been deployed successfully in
the Bitcoin network and extend the Bitcoin protocol by adding a new type of
transaction that hides the origin, destination, and transferred amount from the
public~\cite{sasson2014zerocash}. While a zkSNARK approach has recently been
tested on the Ethereum Byzantium test network, the transaction required two
orders of magnitude more \texttt{gas} to complete, making transaction fees a
potential barrier~\cite{oleary2017private}. Other work, such as
Hawk~\cite{kosba2016hawk}, explores methods for creating privacy-preserving
smart contracts that do not store financial transactions in the clear in order
to retain transactional privacy from the public.

\subsection{Impact of Bugs in Smart Contracts}
\label{sec:bugs}
In contrast to traditional software, smart contracts cannot be directly patched
once deployed. This brings a unique set of challenges and considerations for
designing smart contracts~\cite{porru2017blockchain}. Two common approaches for
updating smart contracts are to either (1) use a self-destruct function that
releases all the internal state of the contract---typically by sending all funds
to a particular address---and then publish a new one, or (2) include a version
flag and a mutable pointer to the address of a new contract after a contract is
deprecated. If a smart contract contains critical flaws, such as logical errors
and lack a self-destruct, assets can be locked in a contract. The most
noticeable example of this was the flaw in The Distributed Autonomous
Organization (The DAO) smart contracts that allowed an attacker to siphon over
\$50M USD out of the \$168M funds invested in the organization. Ultimately, this
required a highly controversial ``hard fork'' to return the state of the
blockchain to a state prior to the hack. Smart contracts may have
vulnerabilities at the language, bytecode, and blockchain levels that open them
to a wide variety of attacks as seen with The DAO, Rubixi, GovernMental, and
others. \cite{atzei2017survey} provides a taxonomy and details of these
exploits.

\section{Conclusion}
\label{sec:conclusion}

IoT applications that automatically perform tasks or exchange assets in behalf
of users pose unique design challenges in terms of: a desire for transparency of
the actions that a device takes on behalf of user; the longevity of these
devices compared to traditional software products; and the common use case of
exchanging digital or physical goods or services, which requires a trusted
arbiter. We make a case for using smart contracts to address these challenges
and take a first look at their trade-offs by designing, implementing, and
evaluating \name, an IoT application for automated gasoline purchases using
machine-to-machine communication. We find that the use of smart contracts
provides transparency, longevity, and allows applications to minimize the need
for trusted third parties---which we estimate can reduce fees paid by a gas
station for a typical transaction by 79\%. However, Ethereum smart contracts
have low transaction throughput (tens of transactions per second) and
95-percentile transaction latency can be on the order of hours, limiting the
types of applications that can be supported.

\para{Acknowledgements}
We would like to thank the anonymous reviewers for their insightful comments and
feedback that helped improve this work. We also gratefully acknowledge the
support of Fujitsu Laboratories Ltd., the Intel/NSF CPS Security grant No.
1505728 and the Stanford Secure Internet of Things Project. We also give a
special thanks to Yan Michalevsky and to the members of the Stanford Information
Networks Group for their discussions and feedback on early versions of this
work.

\IEEEtriggeratref{24}

\bibliographystyle{IEEEtran}
\bibliography{agasp}

\end{document}